\pdfoutput=1

\documentclass[twocolumn]{aastex63}
\usepackage{natbib}
\usepackage{hyperref}
\usepackage{lineno}

\def\deg{\ifmmode^\circ\else$^\circ$\fi}

\shorttitle{intertwined sub-structures in NGC 3324}
\shortauthors{L.~K. Dewangan et al.}

\begin{document}

\title{New insights in the bubble wall of NGC 3324: intertwined sub-structures and a bipolar morphology uncovered by \emph{JWST}}

\correspondingauthor{L.~K. Dewangan}
\email{Email: lokeshd@prl.res.in}

\author[0000-0001-6725-0483]{L.~K. Dewangan}
\affiliation{Astronomy \& Astrophysics Division, Physical Research Laboratory, Navrangpura, Ahmedabad 380009, India.}

\author[0000-0002-7367-9355]{A.~K. Maity}
\affiliation{Astronomy \& Astrophysics Division, Physical Research Laboratory, Navrangpura, Ahmedabad 380009, India.}
\affiliation{Indian Institute of Technology Gandhinagar Palaj, Gandhinagar 382355, India.}

\author[0000-0002-4677-0516]{Y.~D. Mayya}
\affiliation{Instituto Nacional de Astrof\'{i}sica, \'{O}ptica y Electr\'{o}nica, Luis Enrique Erro \# 1, 
Tonantzintla, Puebla, M\'{e}xico C.P. 72840.}

\author[0000-0001-8812-8460]{N.~K.~Bhadari}
\affiliation{Astronomy \& Astrophysics Division, Physical Research Laboratory, Navrangpura, Ahmedabad 380009, India.}
\affiliation{Indian Institute of Technology Gandhinagar Palaj, Gandhinagar 382355, India.}

\author[0000-0002-1920-6055]{Suman Bhattacharyya}
\affiliation{Department of Physics and Electronics, CHRIST (Deemed to be University), Hosur Main Road, Bangalore, 560029, India.}

\author[0000-0001-5731-3057]{Saurabh Sharma}
\affiliation{Aryabhatta Research Institute of Observational Sciences, Manora Peak, Nainital 263002, India.}

\author[0000-0001-8873-1171]{Gourav Banerjee}
\affiliation{Department of Physics and Electronics, CHRIST (Deemed to be University), Hosur Main Road, Bangalore, 560029, India.}

\begin{abstract}
We report the discovery of intertwined/entangled sub-structures towards the bubble wall of NGC 3324 below a physical scale of 4500 AU, which is the sharp edge/ionization front/elongated structure traced at the interface between the H\,{\sc ii} region and the molecular cloud. The sharp edge appears wavy in the {\it Spitzer} 3.6--8.0 $\mu$m images (resolution $\sim$2$''$).  
Star formation signatures have mostly been traced on one side of the ionization front, which lies on the molecular cloud's boundary.
The James Webb Space Telescope's (\emph{JWST}) near- and mid-infrared images (resolution $\sim$0\rlap.{$''$}07 -- 0\rlap.{$''$}7) are employed to resolve the sharp edge, which has a curvature facing the exciting O-type stars. The elongated structures are associated with the 3.3 $\mu$m polycyclic aromatic hydrocarbon (PAH) emission, the 4.05 $\mu$m ionized emission, and the 4.693 $\mu$m H$_{2}$ emission. However, the PAH-emitting structures are depicted between the other two. The H$_{2}$ emission reveals numerous intertwined sub-structures which are not prominently traced in the 3.3 $\mu$m PAH emission. 
The separation between two sub-structures in the H$_{2}$ emission is $\sim$1\rlap.{$''$}1 or 2420 AU. The intertwined sub-structures are traced in the spatial areas associated with the neutral to H$_{2}$ transition zone, suggesting the origin of these structures by ``thin-shell'' instability. Furthermore, an arc-like feature traced in the {\it Spitzer} 3.6--8.0 $\mu$m images is investigated as a bipolar H\,{\sc ii} region (extent $\sim$0.35 pc) at T$_\mathrm{d}$ $\sim$25--28~K using the \emph{JWST} images. A massive star candidate VPHAS-OB1 \#03518 seems to be responsible for the bipolar H\,{\sc ii} region.
\end{abstract}
%
%
\section{Introduction}
\label{sec:intro}
The space-based infrared and sub-millimeter (sub-mm) observations obtained with the {\it Spitzer} and {\it Herschel} 
facilities have revealed bubbles \citep[e.g.,][]{churchwell06,churchwell07} 
and filaments \citep[e.g.,][]{andre10,andre14,Motte+2018} as the major structures in star-forming regions. 
Several previous studies favour the involvement of these structures in the ongoing physical processes of star formation \citep[e.g.,][]{deharveng10,kumar20}. Such observed structures are extensively examined to study the origin of massive OB-stars (M$_*$ $\gtrsim$ 8 M$_{\odot}$) and their feedback processes, which are the open research topics in the study of star formation \citep[e.g.,][]{zinnecker07,tan14,Motte+2018,rosen20}. 
Now, the availability of high-resolution and high sensitivity data from James Webb Space Telescope (\emph{JWST}) provides us a unique opportunity to obtain more insights into physical processes operating in massive star-forming regions \citep[e.g.,][]{pontoppidan22}.
One of the science cases of the \emph{JWST}\footnote[1]{https://webbtelescope.org} is to study the anatomy of Photodissociation Regions (PDRs) surrounding the H\,{\sc ii} regions. The \emph{JWST} images, combined with the sub-mm and radio data sets, can be utilized to probe the molecular zone, the dissociation front, the ionization front, the compressed structure, and the fully ionized flow of gas toward the edge of an H\,{\sc ii} region \citep[see][]{berne22}. 
In this work, we have revisited the publicly available \emph{JWST} near-infrared (NIR) and mid-infrared (MIR) images 
toward the boundary or bubble wall of a nearby young, star-forming region NGC 3324 \citep[or cosmic cliffs NGC 3324; e.g.,][]{pitts19,reiter22}. 
These \emph{JWST} data are part of the public outreach products obtained through the \emph{JWST} Early Release Observations (EROs).
\subsection{Cosmic cliffs NGC 3324}
\label{sec:intxx}
Situated at a distance of 2.2 kpc \citep{goppl22,pontoppidan22,reiter22}, NGC 3324 has a bubble-like appearance \citep[e.g.,][]{smith10,duronea15,reiter22}.
Two O-type stars, HD~92206 and CPD$-$57$^{\circ}$3580 \citep[see][and references therein]{reiter22} have been reported as the major ionizing sources of NGC 3324 (see Figure~\ref{fig1x}a in  Appendix~\ref{sec:posbfield}). 
The selected site in this work (i.e., the bubble rim or NGC 3324 field; see boxes in Figure~\ref{fig1x}a) is a part of NGC 3324, where pillar-like structures have been reported \citep[see Figure 1c in][]{reiter22}. 
The bubble rim in NGC 3324 has the sharpest ionization boundary/front \citep[see][]{smith10,duronea15,pontoppidan22}. 
Using various observational data sets, several studies have been performed toward NGC 3324 \citep[e.g.,][]{smith10,ohlendorf13,preibisch14,duronea15,zeidler16,pitts19}. 
Most recently, using the \emph{JWST} infrared data, \citet{reiter22} reported several outflows including molecular hydrogen objects (MHOs) and Herbig-Haro (HH) flows 
in the NGC 3324 field, and many of these outflows are traced near the ionization front (see their paper for more details). 
It is thought that star formation activity or the presence of a population of actively accreting young stars/young stellar objects (YSOs) in the NGC 3324 field has been affected by feedback from the O-type stars \citep[e.g.,][]{reiter22}.  

Using the \emph{JWST} Near-Infrared Camera \citep[NIRCam;][]{2005SPIE.5904....1R,2012SPIE.8442E..2NB} and Mid-Infrared Instrument \citep[MIRI;][]{2015PASP..127..584R,wright2015_miri} images (resolution $\sim$0\rlap.{$''$}07 -- 0\rlap.{$''$}7), on a small physical scale ($\sim$ 4500~AU), there has been no attempt made to examine the H$_{2}$ emission and the polycyclic aromatic hydrocarbon (PAH) emission toward the ionization front in NGC~3324. 
Additionally, there is no study reported to probe embedded structures (such as bubbles/shells, pillar-like features, and small-scale sub-structures) and their role in the ongoing physical processes in the direction of the NGC 3324 field. 
In this context, we employed ratio and continuum-subtracted maps created from the \emph{JWST} images to study various emission components (i.e., molecular, atomic, and ionized emission) down to physical scale of 4500~AU, allowing us to probe ongoing physical processes in the NGC 3324 field.  

In section~\ref{sec:obser}, we deal with the observational data sets adopted in this paper. 
Our observational findings are presented in Section~\ref{sec:data}. 
In Section~\ref{sec:disc}, we discuss the implications of our outcomes derived toward the NGC 3324 field. 
In Section~\ref{sec:conc}, we provide the conclusions of the present work.
\section{Data sets}
\label{sec:obser}
This paper uses several publicly available data sets at different wavelengths (see Table~\ref{tab1}). 
A large-scale area studied in this paper has an extent of $\sim$8\rlap.{$'$}5 $\times$ 9\rlap.{$'$}8 (central coordinates: $\alpha_{2000}$ = 10$^{h}$36$^{m}$43\rlap.$^{s}$25; $\delta_{2000}$ = $-$58$\degr$37$'$12\rlap.{$''$}82; see a solid box in Figure~\ref{fig1x}a).

We used the Hubble Space Telescope (\emph{HST\footnote[2]{https://archive.stsci.edu/missions-and-data/hst/}}) F658N image (Proposal ID: 10475; PI: Smith, Nathan) taken with the Wide-Field Channel (WFC) of the Advanced Camera for Surveys (ACS). 
In the direction of our target site, the level-3 science ready \emph{JWST\footnote[3]{https://archive.stsci.edu/missions-and-data/jwst}} ERO NIRCam and MIRI images (program ID \#2731; PI: Pontoppidan) were downloaded from the MAST archive. One can also find more details of \emph{JWST} performance in \citet{rigby2023}. More details about these \emph{JWST} data sets are given in \citet{reiter22}. 

The {\it Herschel} dust temperature map\footnote[4]{http://www.astro.cardiff.ac.uk/research/ViaLactea/} (resolution $\sim$12$''$) is explored in this work, which were produced for the {\it EU-funded ViaLactea project} \citep{Molinari10b} using the {\it Herschel} continuum images at 70--500 $\mu$m \citep{Molinari10a} and the Bayesian {\it PPMAP} procedure \citep{marsh15,marsh17}. 
\begin{table*}
 \tiny
\setlength{\tabcolsep}{0.029in}
\centering
\caption{List of observational surveys or data sets utilized in this work.}
\label{tab1}
\begin{tabular}{lcccccr}
\hline 
  Survey/facility  &  Wavelength/      &  Resolution        &  Reference \\   
    &  Frequency/line(s)       &   ($\arcsec$)        &   \\   
\hline
\hline 
Sydney University Molonglo Sky Survey (SUMSS)                 &843 MHz                     & $\sim$45        &\citet{bock99}\\
The Galactic Census of High and Medium-mass Protostars (CHaMP) &  $^{13}$CO (J = 1--0) &   $\sim$37        &\citet{barnes11,barnes18}\\
{\it Herschel} Infrared Galactic Plane Survey (Hi-GAL)                              &70, 160, 250, 350, 500 $\mu$m                     & $\sim$5.8--37.0         &\citet{Molinari10a}\\
\emph{JWST} ERO MIRI F770W, F1130W, F1280W, F1800W imaging facility & 7.7, 11.3, 12.8, 18 $\mu$m                  & $\sim$0.44--0.70          &\citet{2015PASP..127..584R,wright2015_miri}\\ 
{\it Spitzer} Galactic Legacy Infrared Mid-Plane Survey Extraordinaire (GLIMPSE)       &3.6, 4.5, 5.8, 8.0  $\mu$m                   & $\sim$2           &\citet{benjamin03,glimpse1}\\
\emph{JWST} ERO NIRCam Long Wavelength (LW) F335M, F444W, F470N imaging facility & 3.365, 4.421, 4.707 $\mu$m                   
& $\sim$0.17          &\citet{2005SPIE.5904....1R,2012SPIE.8442E..2NB}\\ 
\emph{JWST} ERO NIRCam Short Wavelength (SW) F090W, F187N, F200W imaging facility  & 0.901, 1.874, 1.99 $\mu$m                   
& $\sim$0.07          &\citet{2005SPIE.5904....1R,2012SPIE.8442E..2NB}\\ 
Hubble Space Telescope (HST) ACS/WFC F658N imaging facility                & 6563 \AA + 6583 \AA                     & $\sim$0.067--0.156         &\citet{smith10}\\
\hline          
\end{tabular}
\end{table*}
\section{Results}
\label{sec:data}
A large-scale view of NGC 3324 is presented in Figure~\ref{fig1x}a (see Appendix~\ref{sec:posbfield}). 
In Figures~\ref{fig1x}a--\ref{fig1x}d, the distribution of ionized emission, molecular gas, and YSOs is examined in the direction of cosmic cliffs NGC 3324. 
\subsection{Multi-scale view of cosmic cliffs NGC 3324: intertwined sub-structures}
\label{sec:data2}
Figure~\ref{fig2x}a presents the {\it Spitzer} 5.8 $\mu$m image of the bubble rim.
The visual inspection of the {\it Spitzer} map displays the ionization front (or a sharp edge of the cavity within NGC 3324), and its footprint is also indicated by a dotted curve (in red) in Figure~\ref{fig2x}a. The positions of the previously identified candidate driving sources for the H$_{2}$ outflows \citep[from][]{reiter22} and the YSOs \citep[from][]{kuhn21} are also marked in Figure~\ref{fig2x}a (see also Appendix~\ref{sec:posbfield}).  
The sharp edge has a filamentary appearance, and appears wavy. Massive 870 $\mu$m dust continuum clumps \citep[$>$ 200 M$_{\odot}$;][]{duronea15} 
are distributed toward this elongated and sharp edge. The positions of these clumps are also shown by filled stars in Figure~\ref{fig2x}a. 
From the edge-on view, it appears that we see the sharp or narrow boundary, but from the face-on view, it might just be a sheet. 

A small region labeled as ``sm1'' is indicated by a solid box in Figure~\ref{fig2x}a, where an interesting arc-like feature with dust temperature (T$_\mathrm{d}$) of $\sim$25--28~K is evident. The values of T$_\mathrm{d}$ are obtained from the examination of the publicly available {\it Herschel} dust temperature map (see Appendix~\ref{sec:posbfield} for more details).

Using the {\it Spitzer} 3.6 $\mu$m-band ($\lambda_{eff}$/$\Delta$$\lambda$: 3.55/0.75 $\mu$m) and 4.5 $\mu$m-band ($\lambda_{eff}$/$\Delta$$\lambda$: 4.49/1.0 $\mu$m) images, Figure~\ref{fig2x}b presents a ratio map of 4.5 $\mu$m/3.6 $\mu$m emission for the same area as presented in Figure~\ref{fig2x}a. 
The ratio map allows to separate the features of 4.5 $\mu$m and 3.6 $\mu$m emission as bright and dark areas, respectively.
One can find more details about this ratio map in \citet{dewangan16,dewangan17a}. 
A prominent molecular hydrogen line emission ($\nu$ = 0--0 $S$(9); 4.693 $\mu$m) and a hydrogen recombination line Br$\alpha$ (4.05 $\mu$m) are covered by the {\it Spitzer} 4.5 $\mu$m band. The {\it Spitzer} 3.6 $\mu$m band encompasses a strong PAH emission at 3.3 $\mu$m, which originates from the PDRs. 
The sharp edge or the sharp ionization front is found with the excess 3.6 $\mu$m emission (see dark gray/black areas). 
Several bright areas having the excess 4.5 $\mu$m emission are detected close to the ionization front, where noticeable YSOs and dust clumps are distributed (see Figure~\ref{fig2x}a). 
In the direction of the arc-like feature (see the small region ``sm1'' in Figure~\ref{fig2x}a), the excess 4.5 $\mu$m emission (see bright areas) is traced. 

In Figure~\ref{fig2x}c, we have examined the sharp edge of the cavity within NGC 3324 using the NIRCam image (at \emph{JWST} F335M). 
One can compare different wavelength images of the sharp edge in Figure~\ref{fig2x}.
In particular, the \emph{JWST} F335M image is used to probe sub-structures of the features seen in the {\it Spitzer} 
image (see the sharp edge and the region ``sm1''). One can also find noticeable sub-structures toward wavy structures as appeared in Figures~\ref{fig2x}a and~\ref{fig2x}b. 

The \emph{JWST} NIRCam and MIRI images are presented in Figure~\ref{fig2}. One can note that the coverage of NIRCam and MIRI instruments was taken into consideration while choosing the area shown in Figure~\ref{fig2}, which does not cover the area containing the small region ``sm1'' as presented in Figure~\ref{fig2x}. 
Figure~\ref{fig2}a displays a three-color composite map derived using the \emph{JWST} F470N (in red), F444W (in green), and F335W (in blue) images. 
The F470N~$-$~F444W image has been used to trace the 4.693 $\mu$m continuum subtracted H$_{2}$ emission \citep[e.g.,][]{reiter22}. 
Hence, noticeable outflow signatures depicted in the narrow band F470N ($\lambda_{eff}$/$\Delta$$\lambda$: 4.707/0.051 $\mu$m) image are clearly evident as red color features in Figure~\ref{fig2}a, which have been thoroughly discussed in \citet{reiter22}. The \emph{JWST} NIRCam images (including the F470N $-$ F444W image) reveal the presence of two sub-structures toward the sharp edge of the cavity within NGC 3324, which also seem to overlap each other at many places. 
Hence, such configuration indicates the presence of intertwined sub-structures or double helical-like structures (see arrows in Figure~\ref{fig2}a). 

In Figure~\ref{fig2}b, we display a three-color composite map produced 
using the F1800W (in red), F1130W (in green), and F770W (in blue) images. The locations of previously reported YSOs and candidate driving sources for the H$_{2}$ outflows are also presented in Figure~\ref{fig2}b. 
The intertwined configuration is also evident in the \emph{JWST} MIRI images as seen in the \emph{JWST} NIRCam images. 
In order to further examine the intertwined sub-structures, we processed the \emph{JWST} F1800 image with an 
edge detection algorithm (i.e., ``Edge-DoG'' filter), which uses the technique of Difference of Gaussians filters \citep[e.g.,][]{Assirati2014}. 
The inset of Figure~\ref{fig2}b is a two-color composite map produced using the ``Edge-DoG'' processed \emph{JWST} F1800 image (in red) and 
the \emph{JWST} F1800 image (in turquoise), clearly shows the twisting/coupling of sub-structures (or intertwined sub-structures). 
Although this color composite map is produced for a small area as highlighted by a box in Figure~\ref{fig2}b, similar features can be traced for the entire sharp edge of NGC 3324. 
This is a new finding and its implication is discussed in Section~\ref{sec:disc}.
\subsection{Molecular, neutral, and ionized emission toward intertwined sub-structures}
\label{xxsec:data3}
Figure~\ref{fig8}a displays the \emph{HST} F658N image, which is compared with the \emph{JWST} images (i.e., F444W and (F470N $-$ F444W); see Figures~\ref{fig8}b and~\ref{fig8}c). 
The \emph{HST} F658N filter transmits both H$\alpha$ ($\lambda =$ 6563 \AA) and [NII] ($\lambda =$ 6583 \AA) emission \citep{smith10}. 
The resolution of the \emph{HST} F658N image is equivalent to that of the \emph{JWST} NIRCam SW images, and is better than the spatial resolution of the \emph{JWST} NIRCam LW images (see Table~\ref{tab1}). 

Interestingly, a noticeable elongated bright emission was seen toward the sharp ionization front in the \emph{HST} F658N image. 
We have found that the elongated emission traced in the \emph{HST} F658N image spatially matches with the location of the elongated radio continuum emission as detected 
in the SUMSS 843 MHz radio continuum map (see Figure~\ref{fig1x}c in Appendix~\ref{sec:posbfield}). 
Note that we do not find any sub-structures in the \emph{HST} F658N image (see Figure~\ref{fig8}a).
The twisting of sub-structures associated with the H$_{2}$ emission is presented in the inset of Figure~\ref{fig8}c, which is the \emph{JWST} (F470N $-$ F444W) image exposed to the ``Edge-DoG'' algorithm. In general, the H$_{2}$ molecules get excited by the UV/far-UV (FUV) photons or mechanical heating by shocks (i.e., from outflows, expanding H\,{\sc ii} regions, and stellar winds), which results in the FUV-fluorescence H$_{2}$ emission in a given star-forming region \citep[e.g.,][]{Jo2017}. 
We have examined the extent of the previously reported H$_{2}$ outflows \citep[from][]{reiter22} in the \emph{JWST} (F470N $-$ F444W) image (see Figure~\ref{fig8}c), 
which do not match with our highlighted H$_{2}$ structures. Hence, the filamentary appearance of the H$_{2}$ emission is unlikely to be explained by outflow driven shocks.
Therefore, the H$_{2}$ emission associated with the sub-structures seems to be excited by the UV/FUV heating by ionizing sources of NGC 3324. 

The NIRCam F187N, NIRCam F444W, MIRI F1280W, and MIRI F1800W filters contain the 1.87 $\mu$m Pa-$\alpha$ hydrogen recombination line, the 4.05 $\mu$m Br$\alpha$ hydrogen recombination line, the 12.81 $\mu$m [NeII] line, and the 18.7 $\mu$m [SIII] line, respectively, enabling us to trace the ionized emission. The NIRCam F335M, MIRI F770W, and MIRI F1130W bands cover the 3.3, 7.7, and 11.3 $\mu$m PAH features, respectively \citep{Tielens_2008}. 
Using the \emph{JWST} NIRCam F444W ($\lambda_{eff}$/$\Delta$$\lambda$: 4.421/1.024 $\mu$m) 
and F335M ($\lambda_{eff}$/$\Delta$$\lambda$: 3.365/0.347 $\mu$m) images, we have produced the ratio map of F444W/F335M, which may allow us to probe the 3.3 $\mu$m PAH feature (via dark gray/black areas).  
It is supported with the fact that the \emph{JWST} NIRCam F444W filter does not contain the 3.3 $\mu$m 
PAH feature. Bright areas in this \emph{JWST} ratio map show regions with the prominent 4.05 $\mu$m Br$\alpha$ feature and/or the 4.693 $\mu$m H$_{2}$ emission (see also the interpretation of the {\it Spitzer} ratio map in Section~\ref{sec:data2}). 
We can readily locate the regions with the 4.05 $\mu$m Br$\alpha$ emission in the \emph{JWST} ratio map by taking into account the areas with the 4.693 $\mu$m H$_{2}$ emission detected in the \emph{JWST} F470N $-$ F444W image.

We present Figure~\ref{fig8nw} to trace different emission components towards the sharp ionization front. 
Figure~\ref{fig8nw}a shows a two-color composite map, which is the same as presented in the inset of Figure~\ref{fig2}b.  
In Figure~\ref{fig8nw}b, we display the \emph{JWST} (F470N $-$ F444W) image, which is exposed to the ``Edge-DoG'' algorithm (Figure~\ref{fig8}c). 
Figure~\ref{fig8nw}c presents the ``Edge-DoG'' processed F444W/F335M image, enabling us to depict the PAH emission (dark areas) and the 4.05 $\mu$m Br$\alpha$ emission (bright areas). 
Figures~\ref{fig8nw}a and~\ref{fig8nw}b are shown here only for a comparison purpose. 
The intertwined configuration is not very clearly seen in the ``Edge-DoG'' processed F444W/F335M image, which also seems to reveal the PAH (or neutral) emission and the Br$\alpha$ (or ionized) emission to be in close proximity to each other.  

In Figure~\ref{fig8nw}d, we produce a three-color composite map (``Edge-DoG'' processed F470N$-$F444W image (in red), ``Edge-DoG'' processed F444W/F335M image (in green), and \emph{HST} F658N image (in blue)). In these two ``Edge-DoG'' processed images, to examine the spatial locations of the PAH and H$_{2}$ emission, emission/brightness profiles (not shown here) are also produced 
along a solid line (in yellow) marked in Figure~\ref{fig8nw}d. 
The \emph{HST} F658N bright emission spatially coincides with the 4.05 $\mu$m Br$\alpha$ emission. 
The true boundary of the PDR is traced by H$_{2}$ emission and PAH emission, where the UV light is absorbed. 
More details of these outcomes are discussed in Section~\ref{sec:disc}. 
\subsection{Discovery of a bipolar morphology in the region ``sm1''}
\label{sec:data3z}
\begin{figure*}
\center
\includegraphics[width=\textwidth]{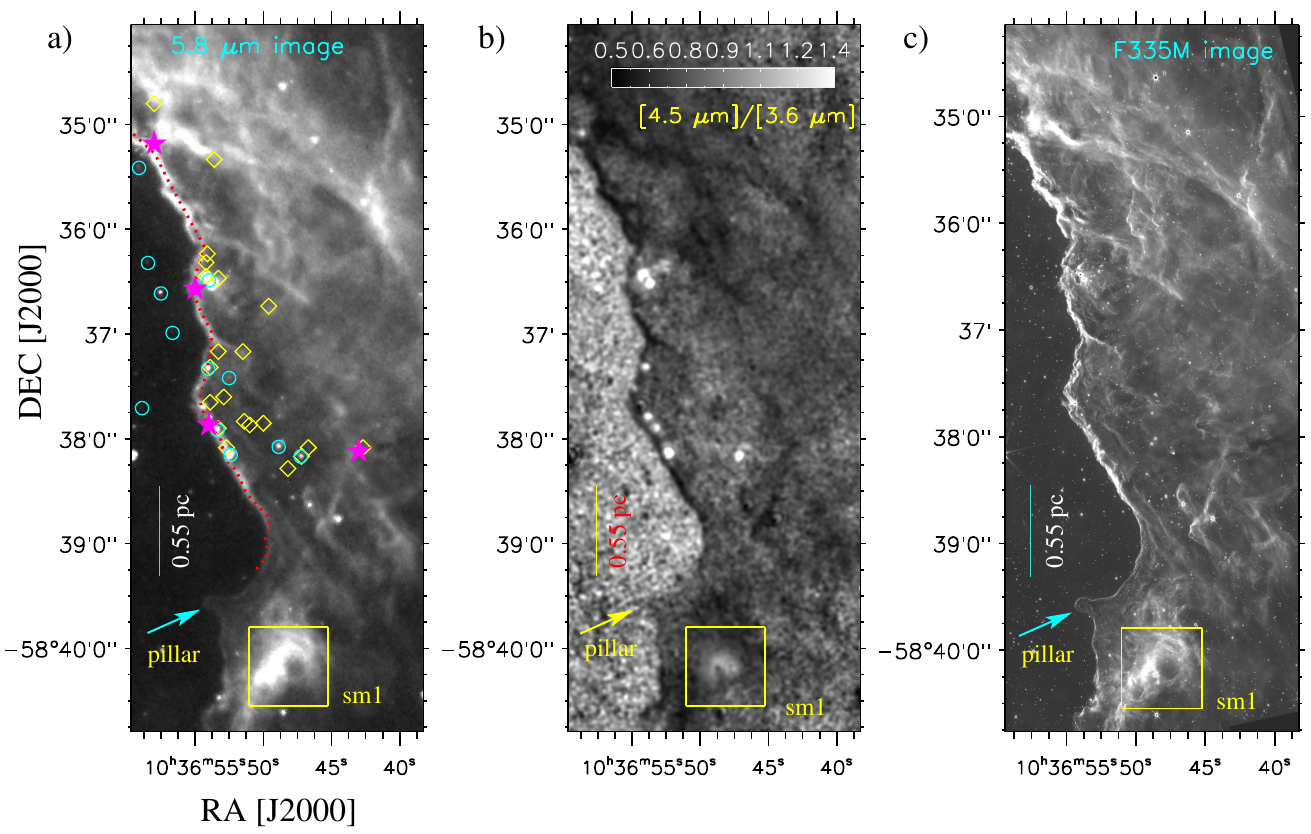}
\caption{a) The panel shows the {\it Spitzer}-GLIMPSE 5.8 $\mu$m image of the bubble rim or cosmic cliffs NGC 3324 (see a dotted-dashed box in Figure~\ref{fig1x}a). 
The area of the image is $\sim$2\rlap.{$'$}79 $\times$ 6\rlap.{$'$}74 (central coordinates: $\alpha_{2000}$ = 10$^{h}$36$^{m}$49\rlap.$^{s}$02; $\delta_{2000}$ = $-$58$\degr$37$'$25\rlap.{$''$}52).
The positions of four clumps traced in the LABOCA 870 $\mu$m continuum map \citep{duronea15} 
are marked by filled stars. Open circles and open diamonds show the locations of YSOs \citep[from][]{kuhn21} and molecular outflows \citep[from][]{reiter22}, respectively. A dotted curve (in red) shows the footprint of the sharp edge. 
b) {\it Spitzer}-GLIMPSE ratio map of 4.5 $\mu$m/3.6 $\mu$m emission. c) The panel displays the \emph{JWST} F335M image. 
In each panel, a small region ``sm1'' is highlighted by a solid box and a scale bar shows a size of 0.55 pc at a distance of 2.2 kpc.} 
\label{fig2x}
\end{figure*}

\begin{figure*}
\center
\includegraphics[width=\textwidth]{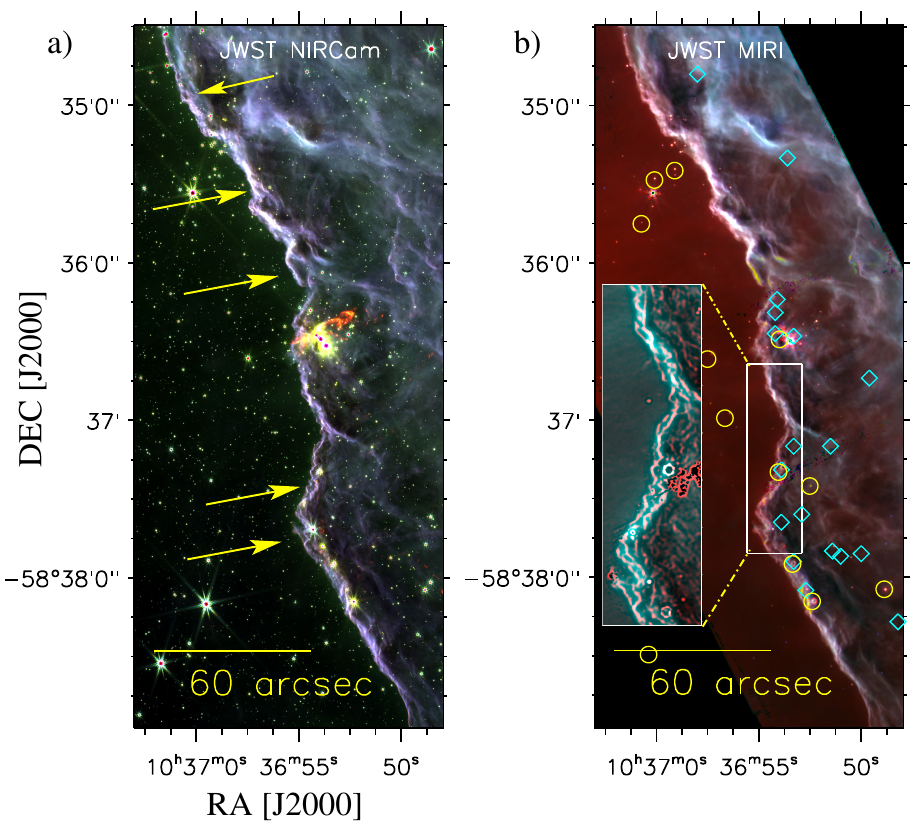}
\caption{a) A three-color composite map produced using the \emph{JWST} F470N (in red), F444W (in green), and F335W (in blue) images (see a dotted box in Figure~\ref{fig1x}a). 
Arrows highlight elongated sub-structures. 
b) A three-color composite map made using the F1800W (in red), F1130W (in green), and F770W (in blue) images.  
Circles and diamonds show the locations of YSOs and molecular outflows, respectively (see Figure~\ref{fig2x}a). 
An inset on the bottom left presents the region in zoomed-in view (see a solid box in Figure~\ref{fig2}b). 
The inset is a two-color composite map produced using the ``Edge-DoG'' processed \emph{JWST} F1800 image (in red) and the \emph{JWST} F1800 image (in turquoise). 
In each panel, a scale bar shows a size of 60$''$ (or 0.64 pc at a distance of 2.2 kpc).} 
\label{fig2}
\end{figure*}

\begin{figure*}
\center
\includegraphics[width=\textwidth]{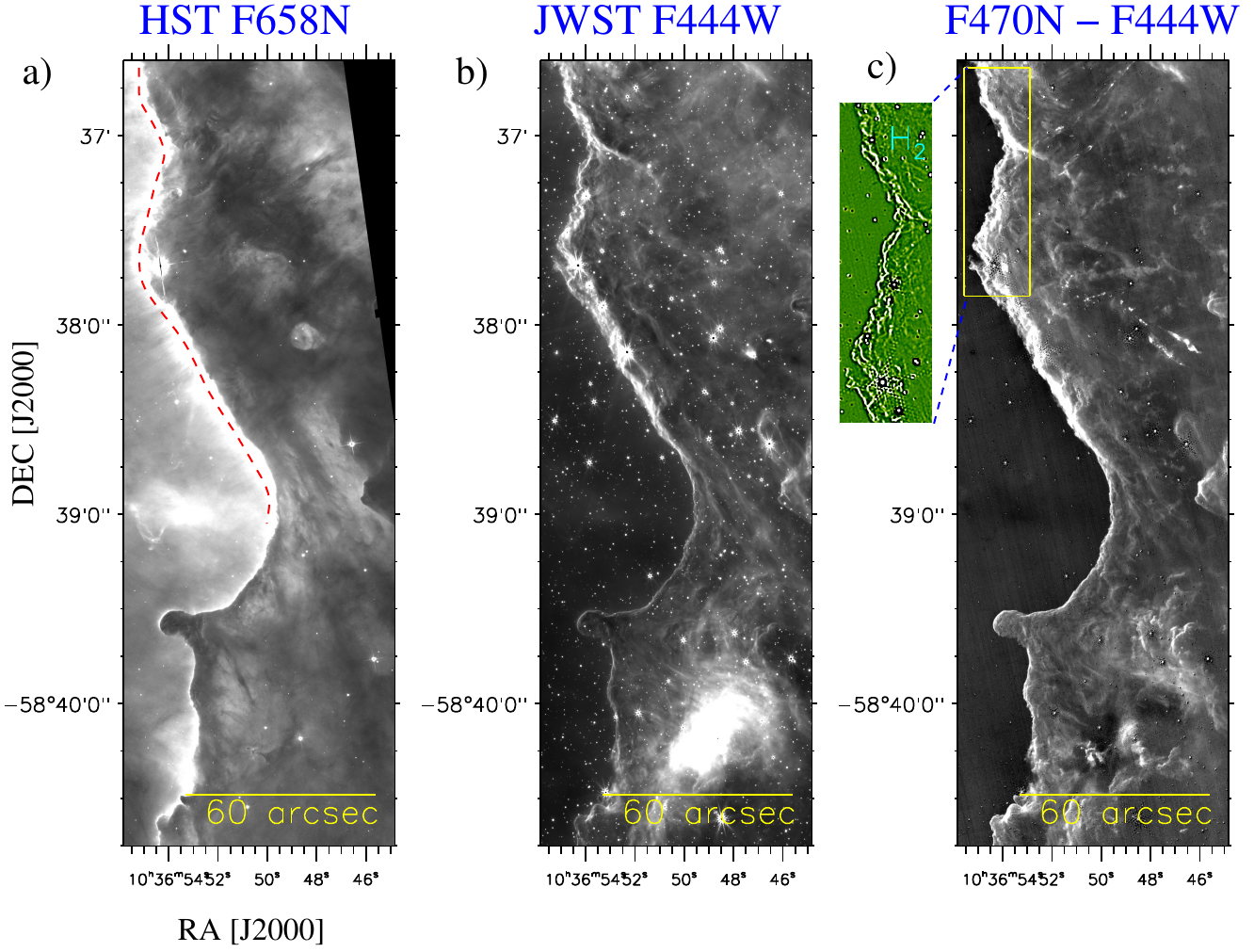}
\caption{a) HST F658N image. A dashed curve (in red) indicates the diffuse emission. 
b) \emph{JWST} F444W image. 
c) F470N$-$F444W (in linear scale). An inset on the top left shows the region in zoomed-in view (see a solid box in Figure~\ref{fig8}c). 
The inset is the ``Edge-DoG'' processed F470N$-$F444W image. In each panel, the scale bar corresponding to 60$''$ (or 0.64 pc at a distance of 2.2 kpc) is shown.}
\label{fig8}
\end{figure*}

\begin{figure*}
\center
\includegraphics[width=\textwidth]{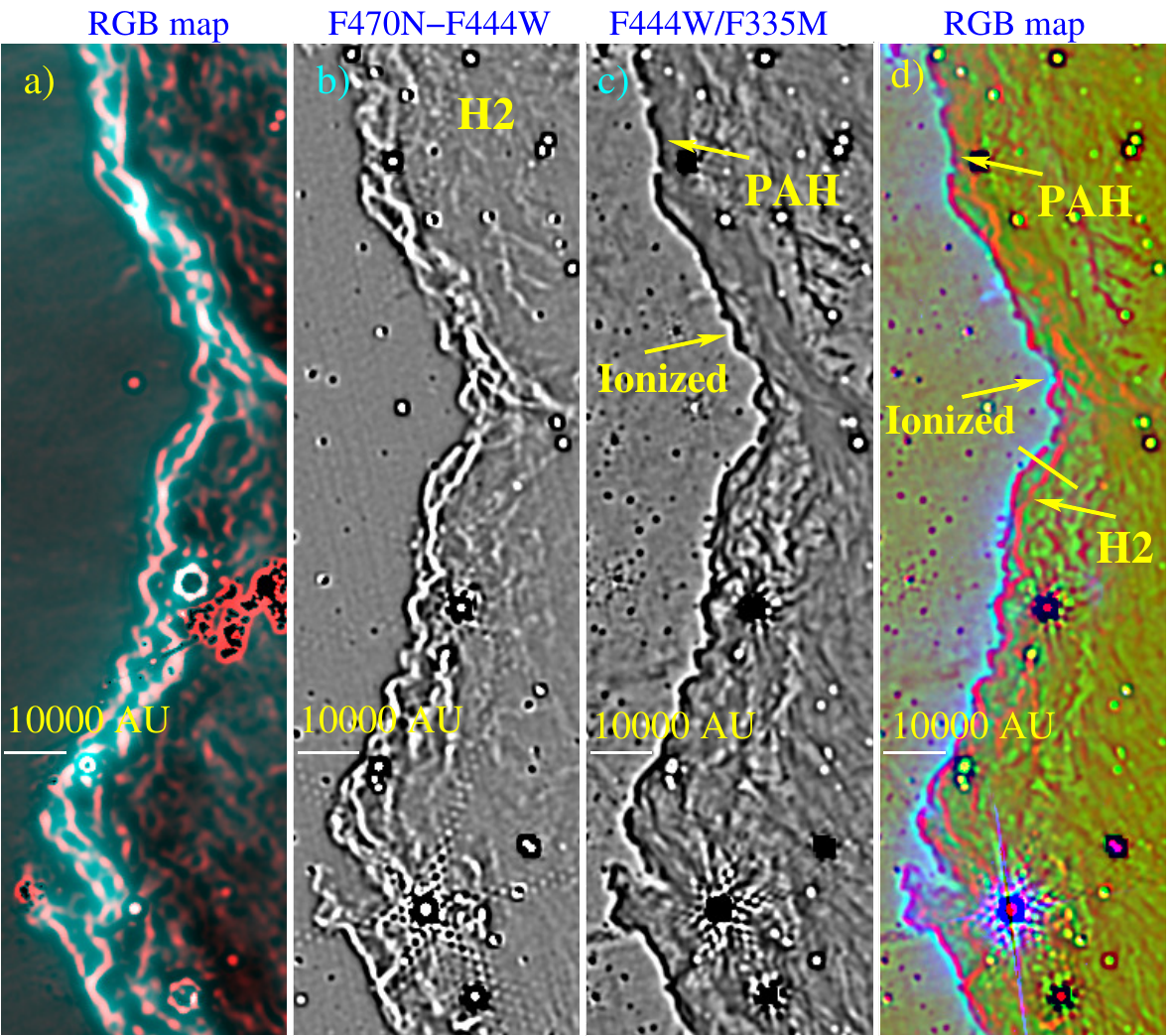}
\caption{
a) Same as the inset shown in Figure~\ref{fig2}b. 
b) Same as the inset shown in Figure~\ref{fig8}c.
c) The panel shows the ``Edge-DoG'' processed F444W/F335M image (see a solid box in Figure~\ref{fig2}b).
d) The panel shows a three-color composite map (``Edge-DoG'' processed F470N$-$F444W image (in red), the ``Edge-DoG'' 
processed F444W/F335M image (in green), and HST F658N image (in blue) images).}
\label{fig8nw}
\end{figure*}

\begin{figure*}
\center
\includegraphics[width=\textwidth]{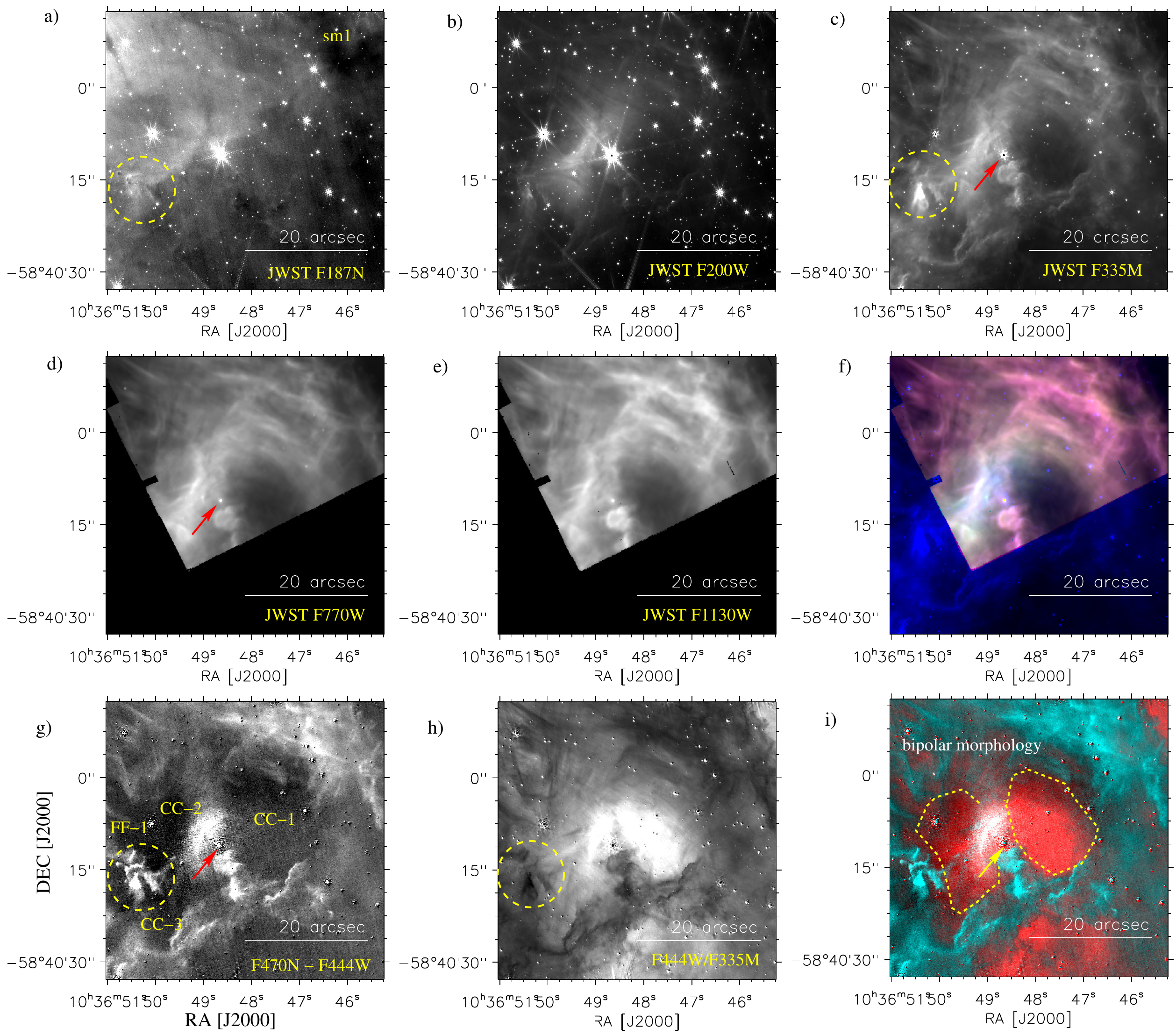}
\caption{Multi-wavelength view of an area ``sm1'' as indicated by a solid box in Figure~\ref{fig2x}a. 
a--e) \emph{JWST} images at F187N, F200W, F335M, F770W, and F1130W are displayed, respectively. 
f) The panel shows a three-color composite map (F1130W (in red), F770W (in green), and F335W (in blue) images in linear scale).
g) F470N$-$F444W (in linear scale).
h) F444W/F335M (in linear scale).
i) The panel presents a two-color composite map (F444W/F335M (red) and F470N$-$F444W (turquoise) images in linear scale).
In panels ``c'', ``d'', and ``g'', an arrow shows the position of an OB-star candidate \citep[UCAC4 157-048728 or VPHAS-OB1 \#03518;][]{mohrsmith17}. 
In panel ``g'', three H$_{2}$ depression regions (i.e, CC-1, CC-2, and CC-3) and one emission structure (FF-1) are labeled.  
In panel ``i'', a bipolar morphology is highlighted, and the position of the OB-star candidate UCAC4 157-048728 is found at its waist. 
In each panel, a scale bar corresponding to 20$''$ (or 0.21 pc at a distance of 2.2 kpc) is drawn.} 
\label{fig3}
\end{figure*}

In the direction of the region ``sm1'', the \emph{JWST} images at different wavelengths are presented in Figures~\ref{fig3}a--\ref{fig3}e, respectively.
As highlighted earlier (see Figure~\ref{fig2x}a), the {\it Spitzer} and {\it Herschel} images of ``sm1'' reveal an arc-like feature with the excess 4.5 $\mu$m emission and the warm dust emission (T$_\mathrm{d}$ $\sim$25--28~K; see Figure~\ref{fig1x}). Using the \emph{JWST} NIRCam and MIRI images, Figure~\ref{fig3}f displays a three-color composite map (F1130W (in red), F770W (in green), and F335W (in blue) images) to examine the locations of the PAH emission. Note that the entire selected area of ``sm1'' is not covered by the \emph{JWST} MIRI images. Figures~\ref{fig3}g and~\ref{fig3}h show the F470N$-$F444W and F444W/F335M images, respectively. In Figure~\ref{fig3}i, a two-color composite map (F444W/F335M (in red) and F470N$-$F444W (in turquoise) images) is presented, where the position of an OB-star candidate \citep[UCAC4 157-048728 or VPHAS-OB1 \#03518; from][]{mohrsmith17} is also indicated by an arrow (see also Figures~\ref{fig3}c and~\ref{fig3}d). 

The presence of H$_{2}$ emission is depicted in the F470N$-$F444W image (see Figure~\ref{fig3}g). 
On the basis of a visual assessment, at least three H$_{2}$ depression regions (i.e, CC-1, CC-2, and CC-3) are identified in Figure~\ref{fig3}g, and are surrounded by the H$_{2}$ emission. Additionally, we have also marked one prominent structure in the \emph{JWST} NIRCam images (see a broken circle in panels of Figure~\ref{fig3}), 
which is found with the intense H$_{2}$ emission (see also Figure~\ref{fig3}g).  
In particular, the region CC-1 is very prominently seen in the longer wavelength images ($>$ 2 $\mu$m), appearing like a bubble feature. 
The distance of the massive star candidate UCAC4 157-048728 or VPHAS-OB1 \#03518 toward ``sm1'' is about 2.3 pc \citep{Jones_2018}, suggesting its association with our target region. The position of UCAC4 157-048728 is not seen at the centre of any H$_{2}$ depression regions (see an arrow in Figure~\ref{fig3}g). 
Based on the published results of the {\it Spitzer} MIR bubbles hosting H\,{\sc ii} regions, the exciting source or massive star is anticipated to be located near the centre of the bubbles \citep[e.g.,][]{churchwell06,churchwell07,deharveng10}.   
Hence, considering the locations of CC-2, CC-3, and the candidate massive star, we avoid proposing CC-1 as a bubble. 
In the direction of our selected area ``sm1'', we do not find any locations of the molecular outflows (see diamonds in Figure~\ref{fig2x}a). Hence, the observed H$_{2}$ emission in Figure~\ref{fig3}g is unlikely to be produced from shocks originating via outflows. 

In Figure~\ref{fig3}h, the F444W/F335M image enables us to trace regions displaying the excess 4.44 $\mu$m emission (see bright areas) 
and/or 3.35 $\mu$m emission (see dark gray/black areas). 
In other words, extended bright areas may show the Br$\alpha$ emission at 4.05 $\mu$m or the ionized emission due to the excess 4.44 $\mu$m emission, while dark gray/black areas may display the PAH emission at 3.3 $\mu$m due to the excess 3.35 $\mu$m emission. 
The H$_{2}$ emission is detected toward the areas having the excess 3.35 $\mu$m emission (see Figures~\ref{fig3}g and~\ref{fig3}h). 
The extended ionized emission traced in the F444W/F335M image appears to be surrounded by the H$_{2}$ emission and the PAH emission (see Figure~\ref{fig3}i).
All the selected H$_{2}$ depression regions (i.e, CC-1, CC-2, and CC-3) seem to be filled with the ionized emission. 

Figure~\ref{fig3}i supports the existence of a bipolar H\,{\sc ii} region or a bipolar morphology (extent $\sim$0.35 pc), and the OB-star candidate appears to be located at the central part of the waist of this bipolar morphology (see dashed curves in Figure~\ref{fig3}i), where we have detected noticeable PAH and H$_{2}$ emissions. 
In known bipolar H\,{\sc ii} regions \citep[see][]{deharveng10}, the location of the exciting source(s) is expected toward the waist of the bipolar structures (e.g., IRAS 17599$-$2148 \citep{dewangan12,dewangan16cc} and W42 \citep{dewangan15}). The structure FF-1 associated with the PAH and H$_{2}$ emissions appears to be located at the edge of the bipolar morphology. 
Locally, the impact of the massive star (UCAC4 157-048728) or UV-fluorescence emission appears to be the cause of the H$_{2}$ emission (see the observed morphology in Figures~\ref{fig3}g and~\ref{fig3}i). 
More discussions on these observed results are presented in Section~\ref{zsec:disc3}. 
\section{Discussion}
\label{sec:disc}
\subsection{Origin of intertwined sub-structures in PDRs}
\label{sec:disc22}
In general, the study of the H\,{\sc ii} regions powered by massive OB stars can allow us to explore the transitional boundary between the neutral/molecular PDR and the fully ionized H\,{\sc ii} region, where temperature and density differences across the boundary can cause evaporative flows and fluid dynamical instabilities \citep[e.g.,][]{hartigan12,goicoechea16}. \citet{goicoechea16} presented a diagram concerning the structure of a strongly UV-irradiated molecular cloud edge, where the ionization front, dissociation front, and the compressed structure are outlined in the PDR near an H\,{\sc ii} region. The PDR initiates at the border of an H\,{\sc ii} region and extends up to the edge of the cold molecular cloud. This predominantly includes neutral and ionized species, resulting from chemical reactions induced by the FUV radiation emitted by the OB stars \citep{goicoechea16}. In PDRs, the H$_2$ emission peaks farther away from the ionizing star and is originated by the UV/FUV radiation or shock heating processes at the shock/dissociation front. 
Furthermore, the Br$\gamma$ (or Br$\alpha$) and H$\alpha$ emission may trace photoevaporative flows from the PDRs \citep[e.g.,][]{carlsten18,wolfire22}. 
Therefore, in general, the different flows (or gas layers) with varying temperatures and densities in the PDR can induce instabilities and give rise to the formation of complex structures. The high-resolution \emph{JWST} images can be employed to directly observe the intricate interfaces of various emissions within the PDR.

The most prominent structure in the bubble wall of NGC 3324 is the sharp edge or the ionization front, which has a curvature that faces the locations of the previously reported O-type stars (i.e., HD 92206 and CPD$-$57$^{\circ}$3580). The observed sharp edge is exposed to the harshest feedback from the powering O-type stars \citep[e.g.,][]{reiter22}. Using the \emph{JWST} NIRCam and MIRI images, the presence of two coupled (or intertwined) sub-structures, below a physical scale of 4500 AU, is investigated toward the sharp ionization front (see Figures~\ref{fig8}c and~\ref{fig8nw}b). 
The elongated structure seen in the 3.3 $\mu$m PAH emission is traced between the ionized emission and the 4.693 $\mu$m H$_{2}$ emission.
The \emph{HST} F658N bright emission spatially coincides with the 4.05 $\mu$m Br$\alpha$ emission. 
In Figure~\ref{fig8nw}d, an offset ($\sim$0\rlap.{$''$}4--1\rlap.{$''$}1 or 880 AU--2420 AU) is evident between the ionized emission 
and the FUV-fluorescence H$_{2}$ emission (see also Section~\ref{xxsec:data3}). The twisting of sub-structures associated with the H$_{2}$ emission is depicted, however such configuration is not clearly traced in the PAH emission toward the higher declination (or northern areas). Based on the intensity profile inspection, the separation between two sub-structures in the H$_{2}$ emission is $\sim$1\rlap.{$''$}1 (or 2420 AU; see a solid yellow line in Figure~\ref{fig8nw}d and also Section~\ref{xxsec:data3}). 
We do not find any previously reported molecular outflows and YSOs towards the elongated structure linked with the PAH emission (see Figure~\ref{fig2}b). 
Interestingly, the spatial areas associated with the neutral/PAH and H$_{2}$ emission in PDRs seem to be resolved in the \emph{JWST} images (see Figure~\ref{fig8nw}d).  

According to \citet{kirsanova23}, the separation between the ionization and dissociation fronts in the S255 H\,{\sc ii} region (d $\sim$2.06 kpc) and the S257 H\,{\sc ii} region (d $\sim$2.5 kpc) is determined to be about 0.3--0.4 pc. In this relation, they used the 2.16 $\mu$m Br$\gamma$ and the 2.12 $\mu$m H$_{2}$ line emission. 
Based on this observational information, they evaluated the applicability of models of a uniform medium and a non-uniform (clumpy) structure of the PDRs. 
They suggested that the derived value of $\sim$0.3--0.4 pc is consistent with a clumpy medium and should be 10--20 times 
lower (i.e., $\sim$3000 AU (or 0.015 pc)--8000 AU (or 0.04 pc)) in a uniform gas density. In our selected target, the H$_{2}$ dissociation front and the H ionization front (or PAH) are not merged.
The transition from PAH to H$_{2}$ is abrupt with sharp edges and with the offset value of $\sim$2420 AU (see Figure~\ref{fig8nw}d). Taking into account these observational findings, numerical simulations of uniform/non-uniform gas density may be carried out to obtain more insights in the bubble wall of NGC 3324. Such exercise, however, is beyond the scope of current work.

It has been proposed that instabilities at the ionization front (or the dissociation front) can be responsible for the irregular shapes of observed H\,{\sc ii} regions including the ``ﬁngers'' and ``elephant trunks'' \citep{giuliani79,garcia96,williams99}. Depending on how thick \citep{williams02} or thin \citep{giuliani79,garcia96} the shell of shocked neutral gas 
is around the ionization front, the onset of different kinds of instability can be inferred \citep[e.g.,][]{henney07}. 

Following the work of \citet{goicoechea16}, it is possible that the force imbalance between thermal (isotropic) and ram pressure (parallel to the flow) is what causes the instability known as the ``thin-shell'' \citep[see also][]{garcia96,williams03}. One can also expect entangled structures in astrophysical plasma on collisional (fluid) scales in the ``thin-shell'' instability \citep{dieckmann15}. Therefore, the intertwined/entangled sub-structures seem to be caused by the ``thin-shell'' instability. 
We further suggest that the intertwined/entangled sub-structures in the PDRs that have been discovered may be considered as one of the direct indicators of instability on the dissociation front. To further gain insights into the observed intertwined configuration, we will need new sub-mm dust continuum and molecular line data with a resolution comparable to the \emph{JWST} images. 

Signposts of star formation including massive stars (i.e., outflows, protostars, 6.7 GHz methanol masers, 22 GHz water masers etc.) are often found at the edges of the {\it Spitzer} bubbles \citep[e.g.,][]{churchwell06,churchwell07} that surround H\,{\sc ii} regions stimulated by massive OB-type stars \citep[e.g.,][]{deharveng10,dewangan16}.
Therefore, the \emph{JWST} images will be very useful to probe ongoing star formation processes toward the edges of the bubbles.
\subsection{Origin of the bipolar morphology seen in the \emph{JWST} images}
\label{zsec:disc3}
As mentioned earlier, a compact and arc-like feature in the small region ``sm1'' is identified using the {\it Spitzer} and {\it Herschel} images (see Section~\ref{sec:data3z}). Diffuse SUMSS 843 MHz radio continuum emission is present towards this selected region (see Figure~\ref{fig1x}). 
Due to the availability of high-resolution \emph{JWST} images, apart from the sharp ionization front, we have also examined this arc-like feature in this paper. In the direction of the arc-like feature, a bipolar morphology (extent $\sim$0.35 pc) in ``sm1'' is investigated using the \emph{JWST} images, and is filled with the ionized emission (see Section~\ref{sec:data3z}). The massive star candidate VPHAS-OB1 \#03518 is found at the waist of the bipolar morphology. In order to infer the impact of VPHAS-OB1 \#03518, we have determined three pressure components (i.e., pressure of an H\,{\sc ii} region ($P_{\mathrm{HII}}$), radiation pressure ($P_{\mathrm{rad}}$), and stellar wind ram pressure (P$_{\mathrm{wind}}$)) driven by the massive OB star following the approach by \citet{dewangan16}.

In the literature, we do not find the exact spectral class of this massive star.  
Therefore, the pressure calculations are performed separately for a B0.5V type star and 
a star of O9.5V type. In general, in case of massive zero age main sequence stars, the value of $P_{\mathrm{HII}}$ exceeds the values of $P_{\mathrm{rad}}$ and $P_{\mathrm{wind}}$ \citep[e.g.,][]{dewangan16}. Note that Wolf-Rayet stars are an exception to this argument, where the P$_{wind}$ value predominates 
over $P_{HII}$ and P$_{rad}$ values \citep[e.g.,][]{lamers99,dewangan16xs,baug19,dewangan22x}.  
Therefore, only the values of $P_{\mathrm{HII}}$ are computed in this paper at $D_{\mathrm{s}}$ = [0.5, 1] pc. The values of $D_{\mathrm{s}}$ are higher than the extent of the bipolar morphology. 
Pressure components driven by O9.5V and B0.5V stars are determined, and are tabulated in Table~\ref{tab2}. 
Table~\ref{tab2} also lists the number of Lyman continuum photons emitted per second (N$_{\mathrm{UV}}$) values for both O9.5V and B0.5V stars \citep[from][]{panagia73}.
The calculation uses the radiative recombination coefficient ($\alpha_{\mathrm{B}}$) = 2.6 $\times$ 10$^{-13}$ cm$^{3}$ s$^{-1}$ at electron temperature ($T_{\mathrm{e}}$) of 10$^{4}$~K. 
In the literature, pressure values (P$_{\mathrm{MC}}$) for typical cool molecular clouds (particle density $\sim$ 10$^{3}$ -- 10$^{4}$ cm$^{-3}$ and temperature $\sim$ 20 K) have been reported to be $\sim$2.8 $\times$ 10$^{-12}$ -- 2.8 $\times$ 10$^{-11}$ dynes cm$^{-2}$ \citep[see Table 7.3 in][]{dyson97}. 
Considering the values of $P_{\mathrm{HII}}$ (for both O9.5V and B0.5V stars) $>$ P$_{\mathrm{MC}}$ (see Table~\ref{tab2}), the impact of a massive OB star at a distance of [0.5, 1] pc seems to be possible. 

Overall, in the selected region ``sm1'' of NGC 3324, the pressure measurements support the impact of the massive star candidate VPHAS-OB1 \#03518 to its surroundings, which is locally responsible for the bipolar morphology as seen in the \emph{JWST} F470N$-$F444W and F444W/F335M images.   
High-resolution sub-mm observations and molecular line data with a resolution comparable to the \emph{JWST} images will be helpful to further explore our proposed arguments in ``sm1''.%
\begin{table*}
\setlength{\tabcolsep}{0.1in}
\centering
\caption{Pressure components driven by O9.5V and B0.5V stars at D$_{s}$ = [0.5, 1] pc (see text for more details). 
The values of N$_{\mathrm{UV}}$ for both the O9.5V and B0.5V stars are also listed.}
\label{tab2}
\begin{tabular}{lccccccccccccc}
\hline 										    	        			      
Spectral type of star &N$_{\mathrm{UV}}$ &    P$_{HII}$ (dynes\, cm$^{-2}$) 	   & P$_{HII}$ (dynes\, cm$^{-2}$)    \\ 
 & (s$^{-1}$) &         at D$_{s}$ = 0.5 pc 	   &   at D$_{s}$ = 1.0 pc    \\ 
\hline 
O9.5V  &1.2 $\times$ 10$^{48}$ &  7.5 $\times$10$^{-10}$  & 2.6 $\times$10$^{-10}$  \\ 
B0.5V  &6.3 $\times$ 10$^{46}$ &  1.7 $\times$10$^{-10}$  & 6.1 $\times$10$^{-11}$   \\ 
\hline          		
\end{tabular}			
\end{table*}			
\section{Summary and Conclusions}
\label{sec:conc}
The current work employs a multi-scale and multi-wavelength approach in the direction of the bubble wall of NGC~3324 (d $\sim$2.2 kpc) to examine ongoing physical processes.
The present paper focuses on the sharp edge/elongated structure or the ionization front in NGC 3324, which is depicted at the interface between the H\,{\sc ii} region and the molecular cloud. 

For the first time, using the \emph{JWST} images at 0.9 --18 $\mu$m, 
we have discovered intertwined/entangled sub-structures or a double helix towards the sharp ionization 
front below a physical scale of 4500 AU. 
The \emph{JWST} F470N$-$F444W image is utilized to study the H$_{2}$ emission at 4.693 $\mu$m, 
while the \emph{JWST} F444W/F335M image has been employed to depict the PAH feature and the ionized emission. 
The elongated structure in the 3.3 $\mu$m PAH emission seems to be seen between the ionized emission and the H$_{2}$ emission.
The presence of the ionized emission is inferred from the Br$\alpha$ (or H$\alpha$) emission. 
We find an offset ($\sim$0\rlap.{$''$}4--1\rlap.{$''$}1 or 880--2420 AU) between the ionized emission and the UV-fluorescence H$_{2}$ emission.

The intertwined configuration is notably evident in the continuum-subtracted H$_{2}$ emission, which is less significant in the PAH emission at 3.3 $\mu$m.
The two sub-structures in the H$_{2}$ emission are separated by at least $\sim$1\rlap.{$''$}1 (or 2420 AU). 
The presence of intertwined structures observed in \emph{JWST} images toward the spatial regions linked to the neutral to H$_{2}$ transition zone suggest that these structures may have originated from the ``thin-shell" instability.

Using the ionized, PAH, and H$_{2}$ emissions traced in the \emph{JWST} images, we have also investigated a bipolar morphology or bipolar H\,{\sc ii} region (extent $\sim$0.35 pc) at T$_\mathrm{d}$ $\sim$25--28~K, which is away from the ionization front and is associated with the elongated molecular/dust cloud.
A previously known massive OB-star candidate (UCAC4 157-048728) is found at the waist of the bipolar morphology, 
and seems to be locally responsible for it. 

To further study the intertwined configuration, sub-mm dust continuum and molecular line data with a resolution comparable to the \emph{JWST} images will be very useful. 
This study also indicates that intertwined sub-structures along the boundaries of the {\it Spitzer} bubbles hosting massive OB-type stars may be commonly detected characteristics that can be investigated with the \emph{JWST} images.
\section*{Acknowledgments}
This work is based on observations made with the NASA/ESA/CSA James Webb Space Telescope. The data were obtained from the Mikulski Archive for Space Telescopes at the Space Telescope Science Institute, which is operated by the Association of Universities for Research in Astronomy, Inc., under NASA contract NAS 5-03127 for \emph{JWST}. These observations are associated with program \#2731. The specific observations analyzed can be accessed via \dataset[10.17909/stwg-r373]{http://dx.doi.org/10.17909/stwg-r373}.

The authors thank the anonymous referee for helpful comments and suggestions that improved the manuscript.  
The research work at Physical Research Laboratory is funded by the Department of Space, Government of India. Y.D.M. thanks CONACyT (Mexico) for the research grant CB-A1-S-25070(YDM). This work is based [in part] on observations made with the {\it Spitzer} Space Telescope, which is operated by the Jet Propulsion Laboratory, California Institute of Technology under a contract with NASA. 
This research has made use of the VizieR catalogue access tool, CDS,
 Strasbourg, France (DOI : 10.26093/cds/vizier). The original description 
 of the VizieR service was published in 2000, A\&AS 143, 23.

\appendix
\restartappendixnumbering
\section{Multi-wavelength view of cosmic cliffs NGC 3324}
\label{sec:posbfield} 
\begin{figure*}
\center
\includegraphics[width=\textwidth]{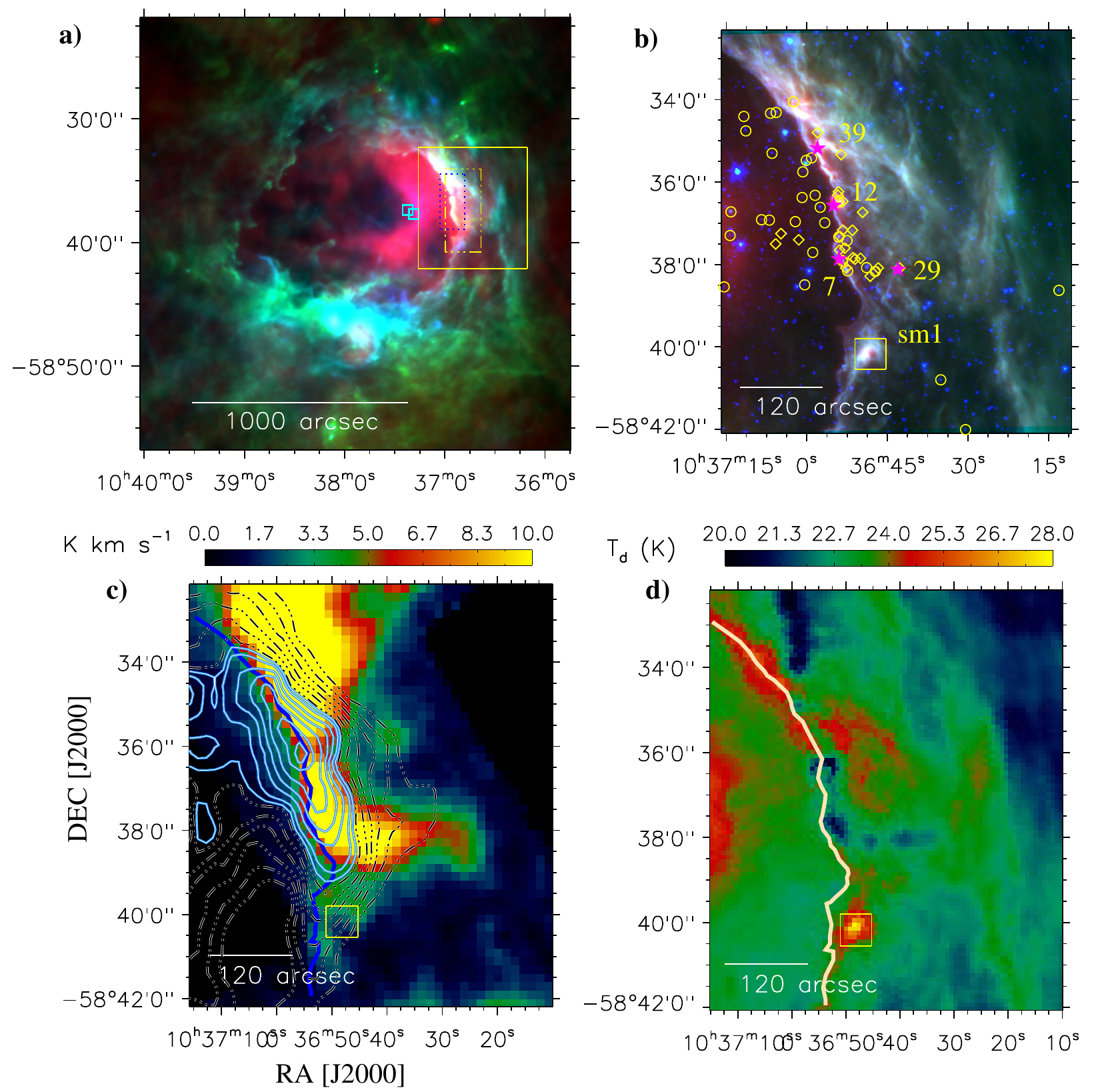}
\caption{a) Large scale view of NGC 3324 using a three-color composite map (SUMSS 843 MHz (in red), {\it Herschel} 160 $\mu$m (in green), and {\it Herschel} 70 $\mu$m (in blue) images).
Squares show the locations of massive O-type stars. Boxes indicate areas which are examined in this work. 
A solid box emcomapsses an area presented in Figures~\ref{fig1x}b,~\ref{fig1x}c, and~\ref{fig1x}d. b) The panel shows a three-color composite map ({\it Herschel} 70 $\mu$m (in red), {\it Spitzer} 8.0 $\mu$m (in green), and {\it Spitzer} 3.6 $\mu$m (in blue) images in linear scale). 
The area of the image is $\sim$8\rlap.{$'$}5 $\times$ 9\rlap.{$'$}8 (central coordinates: $\alpha_{2000}$ = 10$^{h}$36$^{m}$43\rlap.$^{s}$25; $\delta_{2000}$ = $-$58$\degr$37$'$12\rlap.{$''$}82).
All the symbols are the same as presented in Figure~\ref{fig2x}a. The dust continuum clumps highlighted by stars are also labeled. c) Overlay of the SUMSS 843 MHz radio continuum contours on the CHaMP $^{13}$CO moment-0 map at [$-$24.3, $-$19.9] km s$^{-1}$. 
The SUMSS 843 MHz radio continuum contours (see dotted-dashed contours) are plotted with the levels of 46.5, 62.3, 78, 94, 109, and 125 mJy beam$^{-1}$. 
The SUMSS contours (see solid contours) are presented with the levels of 132, 141, 157, 172, 188, 204, and 212 mJy beam$^{-1}$. d) {\it Herschel} dust temperature map. 
In panels ``c'' and ``d'', a thick curve shows the footprint of the sharp edge as seen in Figure~\ref{fig1x}b. 
In panels ``b'', ``c'' and ``d'', a small region (i.e., ``sm1'') is indicated by a small yellow box. 
Scale bars are shown (at 1000$''$ (or 10.7 pc at a distance of 2.2 kpc) and 120$''$ (or 1.28 pc)).}
\label{fig1x}
\end{figure*}

The morphology of NGC 3324 is presented in Figure~\ref{fig1x}a, which is a three-color composite map 
(SUMSS 843 MHz (in red), {\it Herschel} 160 $\mu$m (in green), and {\it Herschel} 70 $\mu$m (in blue) images). 
The locations of massive O-type stars (i.e., HD~92206 and CPD$-$57$^{\circ}$3580) are also marked by squares in Figure~\ref{fig1x}a. 
The free-free emission tracing the ionized nebula in the SUMSS 843 MHz radio continuum map is distributed within 
the extended bubble NGC 3324. An area highlighted by a solid box in Figure~\ref{fig1x}a is presented in Figure~\ref{fig1x}b, 
which shows a three-color composite map produced using the {\it Herschel} 70.0 $\mu$m far-infrared image (in red), 
{\it Spitzer} 8.0 $\mu$m MIR image (in green), and {\it Spitzer} 3.6 $\mu$m NIR image (in blue). 
In Figure~\ref{fig1x}b, the positions of YSOs, 870 $\mu$m dust continuum clumps \citep[at d $\sim$2.5 kpc; IDs: 7 (mass $\sim$530 M$_{\odot}$), 12 (mass $\sim$390 M$_{\odot}$), 29 (mass $\sim$390 M$_{\odot}$), and 39 (mass $\sim$220 M$_{\odot}$); see][for more details]{duronea15}, and molecular outflows are also highlighted by open circles, filled stars, and open diamonds, respectively. 
This paper is mainly focused on the sharp edge of the cavity within NGC 3324 (or the ionization front) and the arc-like feature toward the small region ``sm1'', which are evident in 
the {\it Spitzer} and {\it Herschel} images. 
Massive dust continuum clumps \citep[mass range $\sim$220--530 M$_{\odot}$;][]{duronea15} have been reported toward the sharp edge.

In the direction of cosmic cliffs NGC 3324, Figure~\ref{fig1x}c displays the integrated intensity (i.e., moment-0) map of CHaMP $^{13}$CO(J =1$-$0) over a velocity range of [$-$24.3, $-$19.9] km s$^{-1}$, revealing an elongated morphology. The moment-0 map is also overlaid with the SUMSS 843 MHz radio continuum contours. 
The radio continuum map favours the presence of an elongated feature (extent $\sim$7\rlap.{$'$}5) having an aspect ratio (i.e., length/width) of $\sim$3, 
which seems to be located toward the sharp ionization front. 
Previously, \citet{duronea15} explored these two data sets toward NGC 3324. 
However, they did not discuss any elongated ionized feature. In the literature, we find that the energetic feedback from massive stars has been proposed to explain the existence of the elongated ionized features by swept-up ionized gas or dissipated turbulence \citep[e.g.,][]{emig22}. Hence, such ionized features are expected to be far from the exciting massive stars in the OB-association/ OB-star complex \citep[e.g.,][]{karr03,pon14,emig22}. Based on the spatial location of the elongated ionized feature with respect to the ionization front and the O-type stars, we suggest that the feedback from the massive O-type stars is likely the reason for the existence of the elongated ionized feature. 

In Figure~\ref{fig1x}d, we present the {\it Herschel} dust temperature map. 
Figures~\ref{fig1x}c and~\ref{fig1x}d covers the same area as displayed in Figure~\ref{fig1x}b. 
The footprint of the ionization front is also marked by a solid curve in Figures~\ref{fig1x}c and~\ref{fig1x}d. On the basis of Figures~\ref{fig1x}c and~\ref{fig1x}d, we can study the spatial distribution of the molecular gas, dust temperature, and ionized emission with respect to the ionization front. At the molecular cloud's edge, the ionization front is seen, where variations of the dust temperature (T$_\mathrm{d}$ $\sim$20--28~K) are evident. Overall, the sharp edge appears to be situated at the molecular gas/H\,{\sc ii} region interface. 

\bibliographystyle{aasjournal}
\bibliography{reference}{}


\end{document}